\documentclass[8pt,fleqn,twoside]{article}
\usepackage{float}
\usepackage{amsmath}
\usepackage{amssymb}
\usepackage{graphicx}
\newcommand{\ebnercite}[1]
 {$^\text{\scriptsize \cite{#1}}$}
\newcommand{\ebnerboldmath}[1]{\mbox{\boldmath $#1$}}
\newcommand{\ebeq}[1]{(#1)}
\begin{document}
\thispagestyle{empty}

{\bf How Hilbert has found the Einstein Equations before Einstein
and forgeries of Hilbert's page proofs}

{\bf Dieter W. Ebner\\}
{\footnotesize
 Physics Department, University of Konstanz, DE-78457 Konstanz\\
 e-mail: Dieter.Ebner@uni-konstanz.de\\
}

{\bf Abstract}: A succinct chronology is given around Nov 1915,
when the explicit field equations of General Relativity have been found.
Evidence, unearthed by D.Wuensch, that a decisive document of
Hilbert has been mutilated in recent years with the intention to
distort the historical truth is reviewed and discussed. The
procedure how Hilbert has found before Einstein the correct
equations ``easily without calculation" by invariant-theoretical
arguments is identified for the first time. However, Hilbert has
based his derivation on an incorrect or at least not yet formally
proved invariant theoretical fact.

{\footnotesize {\em Key words:} History of General Relativity, Field Equations of
General Relativity
}
\section{Introduction}
At Nov 11, 1915 Einstein still submits incorrect field-equations
\ebeq{2} of General Relativity, lacking the trace term, to
\emph{Preu{\ss}ische Akademie der Wissenschaften zu Berlin}%
\ebnercite{EinsteinNov4und11}%
.
At Nov 20, for the first time, David Hilbert submits
the correct equations \ebeq{5c} to \emph{Gesellschaft der Wissenschaften
zu G\"ottingen}%
\ebnercite{PageProofs}%
.
Einstein, 5 days later, now also submits the correct equations \ebeq{7} at Berlin%
\ebnercite{EinsteinNov25}%
, but without citing Hilbert, although with a postcard%
\ebnercite{EinsteinBriefe}
dated Nov 18, Einstein acknowledges receipt of a postcard from Hilbert
(probably dated Nov 16)
containing field equations for General Relativity. Hilbert's postcard is lost.

Most physicist, including those working on General Relativity,
either have no definite opinion or they believe that General
Relativity was the creation of Einstein alone. Only a few,
interested in history of science, believed that Hilbert has first
published the correct equations, and Einstein 5 days later, either
independently or through the influence of Hilbert, arrived also at
the correct equations.

In 1997 this opinion was challenged by L.Corry, J.Renn and
J.Stachel (CRS). After having found in a G\"ottingen archive the page
proofs%
\ebnercite{PageProofs}
of Hilbert's Nov 20 submission, which he has received from the
printery at Dec 6, they publish in {\em Science} an article%
\ebnercite{Science}
stating: {\em nowhere} in the whole page proofs the correct
equations can be found. Thus they revert the opinion held thus far
by historians, by insinuating Hilbert has copied from Einstein,
and not Einstein from Hilbert, because
Hilbert had added the
explicit equations \ebeq{5c} with the correct trace term only after Dec
6, appearing in his published version%
\ebnercite{HilbertPublished}
at Mar 31, 1916 only. Hilbert has made other considerable changes
for his published version without altering the initial date Nov 20
of submission or adding a date of revision.

After 1997, a tremendous fact was unearthed, turning the
Einstein/Hilbert priority dispute from a purely academic
controversy among historians into a criminal story: In 2002
Winterberg discovers that about one third of a page in Hilbert's page
proofs have been cut off, and Winterberg was the first to publish%
\ebnercite{Winterberg}
the opinion the cut-off is the deed of a forger with the intention to distort the
historical truth in favour of Einstein.
(For a list of earlier mentioning of the cut-off see Sommer%
\ebnercite{Sommer}%
.)

CRS defend with vehemence their opinion%
\ebnercite{Vehemence}
Hilbert himself had made the cut-off. However, the most miraculous
item in this case would then be the fact, that CRS
did not mention the cut-off in their {\em Science} article, though their
main statement was that {\em nowhere} in the whole page proofs the
explicit equations can be found.

The refereeing process in {\em Science} is very strict. Therefore we
conclude that their referees have been in possession of a copy of
the intact page proofs. Otherwise the referees had observed the
cut-off and they had not passed the article in the present form.
Similarly we conclude that indeed the intact page proofs did not
contain the explicit field equations.

For historians it was a great triumph%
\ebnercite{Triumph}
to have reverted the opinion
about such a milestone in the development of physics as is the discovery of the
correct explicit field equations of General Relativity. However, their triumph was shattered
by the early argument Hilbert was the first to have identified the Ricci scalar $R$
as the correct Lagrangian density and the elaboration by well known variational
procedures were only a trivial exercise, thus Hilbert still deserved the priority.
Thus a forger went out and made the cut-off in Hilbert's page proofs. In the
remaining page proofs it can no longer be seen that Hilbert's $K$ is indeed the
`invariant stemming from Riemann's tensor', i.e. the Ricci-scalar $R$, and
Hilbert could have added that statement only for his published version after
he has seen the correct field equations in Einstein's publications.

Hilbert's postcard (dated Nov 16 $\pm$ 1 day) in which he has
communicated the correct explicit field equations to Einstein is lost,
which is curious since more trivial postcards before and later
are preserved. Only from Einstein's reactions we can safely infer that
Hilbert's Nov 16 postcard indeed contained the correct field equations.

Hilbert's case was also weak because Hilbert did not give detailed
explanations how he has found the equations, but says they follow
`easily without calculation' only mentioning invariant theoretical
arguments. A straightforward elaboration of the variational
procedure, without using any tricks discovered only later, though
certainly possible, would require several days of calculation. I
doubt if anyone has really done this `great calculation', as I
call it in the following, and it seems at least nobody has
published it.

In the published version of his page proofs, Hilbert only gives the
following explanations:
\begin{gather*}
\tag{1}
\boxed
{
\begin{matrix}
\text{The first term on the left hand side of $[$our Eq.\ebeq{5c}$]$
  follows easily without calculation}\\
\text{from the fact that $K_{\mu\nu}$, except $g_{\mu\nu}$, is the
  only second rank tensor, and $K$ is the only}\\
\text{invariant which can be formed from $g^{\mu\nu}$ and
  its first and second derivatives $g_k^{\mu\nu}, g_{kl}^{\mu\nu}$.}\\
\end{matrix}
}
\end{gather*}
(Note that Hilbert writes $K$ for the Ricci scalar $R$, $K_{\mu\nu}$ for
the Ricci tensor $R_{\mu\nu}$, and he adds additional lower indices for
partial derivatives with respect to the coordinates.)

I have pondered about Hilbert's `easy' method, since I was
dissatisfied by explanations given by others%
\ebnercite{Vermeil}%
, who have proposed simple but tricky procedures, and probably are
methods discovered only later. Coincidentally I have ordered
further copies of Hilbert's private notes from the G\"ottingen archive%
\ebnercite{OrderFromArchive}%
, and I was puzzled about Hilbert mentioning, among general
formulae he has obtained for Riemannian geometry, a `special case'
which seemed completely absurd for me until I realized that this
`special case' is sufficient to derive the Einstein Equations. For
a skilled theorist, as Hilbert was, that calculation required
perhaps 5-10 minutes, thus he described it as `easily without
calculation' meaning avoiding the straightforward but lengthy
`great calculation', which would require several days. Hilbert's
mentioning that `special case' in his private notices is a hint he
actually proceeded along these lines. A lot of other special cases
would do also. Therefore Hilbert did not specify one in his
published version. In the following subsection 6.3 we give an even
more elegant derivation, deserving the qualification `easily
without calculation'. It is based on a tacit assumption, which
Hilbert was allowed using his famous power of mathematical
intuition.

Unfortunately for Hilbert, the fact mentioned in \ebeq{1}, as it
is stated, is incorrect, even if one assumes that with `tensor'
Hilbert means `symmetric tensor', which perhaps was his
terminology. Obviously Hilbert did not know the results of Haskins%
\ebnercite{Haskins}
and Zorawski%
\ebnercite{Zorawski1}%
, though in 1908 the ``\emph{Leipziger Berichte}" were available at
G\"ottingen. According to their results, see the following section
\ebeq{7.1}, for $n=4$, including the Ricci scalar and Ricci tensor,
there are 14 independent invariants and three tensors. Fortunately
for Hilbert, the fact mentioned in \ebeq{1} is correct if invariant
and tensor are required to be linear in the second derivatives
$g_{kl}^{\mu\nu}$. That modified statement \ebeq{1} suffices for
Hilbert's derivation.

However, that
modified fact \ebeq{1} was not known in 1915, and so Hilbert must have been
aware his derivation of the explicit field equations is based on a
shaky invariant theoretical fact, at least lacking a formal proof.
This may explain why Hilbert did not include them in his page
proofs. Also because for him, as a mathematician, explicit field
equations did not seem to be of utmost importance. Only after he
has seen them in Einstein's Nov 25 publication (without acknowledgment
of his private Nov 16 postcard communication) he has added them
in a hurry, together with the insufficient explanation \ebeq{1}.

In his republication%
\ebnercite{Hilbert1924}
1924, i.e. almost ten years later, Hilbert writes in
the introduction:
{\em ``The following is essentially a reprint of my older
communications ... with only slight editorial changes
and rearrangements, which should improve its comprehension." }
Despite that statement, he abandons his previous derivation based on \ebeq{1}
completely and switches to the standard derivation,
which in the meantime was found by others,
using Riemannian normal coordinates.
In the meantime Vermeil%
\ebnercite{Vermeil2}%
, Weyl%
\ebnercite{Weyl2}
and Cartan%
\ebnercite{Cartan}
have given independent formal
proofs of Hilbert's modified fact.
I do not see any reason for a
republication, except Hilbert has realized his 1915 sketch of a derivation
of the field equations based on \ebeq{1} was incomplete. His introductory words should hide
his own derivation was a guess only and not a formal proof.

\section{A Chronology%
\ebnercite{prioritydispute}%
}
Since {\bf 1907} Einstein was working toward a unification of
special relativity with gravitation. He had the ingenious idea
that gravitation has something to do with curvature of space-time,
thus unifying geometry and gravitation, or geometrizing the
gravitational force. The metrical tensor $g_{ab}$ (depending upon
chosen generalized coordinates $x^k$) is the formulation of the
metric {\em and} of the gravitational field, replacing Cartesian
coordinates and Newton's gravitational potential. Masses
(equivalent to energy according to $E=mc^2$) are the source of
curvature.

{\bf 1913} M. Grossmann%
\ebnercite{Grossmann}%
, for the first time, proposed the Ricci-tensor
$R_{ab}$ as a covariant candidate
expressing curvature in a gravitation theory as intended by Einstein.
Unfortunately, Grossmann erroneously concluded that equations
with the Ricci tensor cannot reproduce Newton's theory of gravitation
in the limit of weak gravitational fields. According to the published
version%
\ebnercite{Grossmann}%
, the error is due to Grossmann. However, in his
Nov 4 paper (see below), Einstein takes the full
responsibility for the error.

{\bf 1914 and the first half of 1915}: Tragically, based on this
misjudgement by Grossmann, Einstein publishes several papers
stating the gravitational field equations in the form of the
following Eq. \ebeq{2}, but with the left-hand side not the
Ricci-tensor, but several curious mathematical expressions, which
are not generally covariant but only covariant under
linear transformations.\\
(On the right hand side of \ebeq{2}, $\kappa$ is the constant of gravitation
in certain units, and $T_{ab}$ is the energy-momentum tensor of matter or
radiation, producing the curvature. $T_{00}$ is the
mass density. The other components are pressure, stress,
and mass flux density, which also produce curvature.)

General covariance of a field equation such as \ebeq{2}
is mandatory for
General Relativity, which is a theory of the metrical field. General covariance
means that arbitrary coordinates might be chosen and the field equations such as
\ebeq{2} have the same form (covariance = form invariance), or in other
words \ebeq{2} should be valid in any coordinate system.
Special coordinates can be defined using the metric.
Thus special
coordinates bear the danger that they restrict the generality of
the gravitational field.

{\bf June 28 - July 5, 1915} By Hilbert's invitation,
Einstein gives six two-hour lectures about General Relativity
in G\"ottingen.

{\bf Nov. 4, 1915}
For the first time, Einstein%
\ebnercite{EinsteinNov4und11}
postulates the field equations in the form \ebeq{2}
with the left-hand side the Ricci-tensor. But unnecessarily,
he uses a special expression for the
Ricci tensor, which is valid only in special
coordinates satisfying \ebeq{4}, where $g$
is the determinant of the metrical tensor, so {\em manifest} general
covariance is not yet reached.\\

The most important
progress in that paper was that Einstein proved that \ebeq{2}
has Newton's gravity as a limit, thus correcting
Grossmann's error, which had cost Einstein almost
3 years of futile work. Einstein writes%
\ebnercite{EinsteinNov4und11}%
: {\em
``Thus I regained general covariance, a requirement
which I had abandoned with a heavy heart
three years ago, while working with my friend Grossmann."}

At the same time, Einstein freed himself from another quarrel he
has had against his theory, namely his earlier observation that
all equations of the form \ebeq{2}, with the left hand side having some
(non-trivial, i.e. local) covariance, violate causality: Knowledge
of the metrical tensor field in the past cannot predict it
uniquely in the future with the help of \ebeq{2}. Einstein is now
recognizing that this is because of covariance: Coordinates have
no objective physical meaning and our free choice in the future
(permitted by general covariance) cannot be predicted by a
physical law. Nevertheless, coordinates are necessary to formulate
the metrical-gravitational field, namely with the help of a
metrical tensor $g_{ab}$. Therefore \ebeq{2} must and can be
supplemented by 4 additional conditions, nowadays called
gauge-conditions, specifying the choice of coordinates in the
future, thus restoring causality of the field equations of General
Relativity.

{\bf Nov 7, 1915} Einstein sends Hilbert the page proofs of the
above Nov 4 paper.

{\bf Nov 11, 1915}
For the first time, Einstein%
\ebnercite{EinsteinNov4und11}
proposes (manifestly) fully covariant field
equations for the gravitational field:
\begin{equation*}
\tag{2}
R_{ab}= -\kappa T_{ab}
\end{equation*}
where $R_{ab}$ is now the Ricci tensor formulated in arbitrary coordinates.
(The minus sign on the right-hand
side of \ebeq{2} is conventional,
depending on a conventional minus sign in the definition of the Ricci-tensor.)
A milestone in the development of General Relativity was hereby achieved:
\ebeq{2} has Newton's theory of gravitation as a limit, though with $\kappa$
half the value as given by \ebeq{8}, {\em and} \ebeq{2} are (manifestly) generally
covariant field equations, {\em and} most spectacularly (see Einstein's Nov 18 paper)
\ebeq{2} explains Mercury's perihelion precession, {\em and} \ebeq{2} explains the
deflection of light as (allegedly) observed at the 1919 eclipse of the sun by Eddington
and Fr\"ohlich.

Nevertheless Einstein was deeply dissatisfied with \ebeq{2}.
As is now also the accepted view, Einstein was
convinced
that to every physical equation in special relativity (i.e. in the absence
of a gravitational field, e.g. conservation of energy and momentum, Maxwell's equations, etc)
there corresponds an analogous equation in General Relativity (i.e. in the
presence of a gravitational field) whereby the partial derivatives with respect to the
coordinates are replaced by covariant derivatives. Nowadays this is called the
{\em principle of
minimal coupling}, and is the accepted method for postulating Maxwell's equations in
General Relativity. Conservation of energy and momentum in flat Minkowski space
of special relativity thus leads
to the equation
\begin{equation*}
\tag{3}
T^b_{a;b}=0
\end{equation*}
where $;$ denotes covariant derivative. Einstein observed that \ebeq{2}
does not satisfy \ebeq{3} except in the case the trace $T=T_a^a$
vanishes. As we know today, the discrepancy between \ebeq{2} and \ebeq{3} is
serious because applied to a point particle (e.g. the earth), \ebeq{3}
is equivalent to geodesic motion. Thus the empirical success of
\ebeq{2} is spoilt, because being based on the assumption of geodesic
motion in the Schwarzschild metric.

{\bf Nov 12, 1915} Einstein
writes to Hilbert: {\em ``Meanwhile, the problem has been brought
one step forward. Namely, the postulate
\begin{equation*}
\tag{4}
\sqrt{-g} = 1
\end{equation*}
enforces general covariance."} In modern terms, this means that
the gauge-condition \ebeq{4} (nowadays called Einstein gauge) is a {\em
permitted} gauge, i.e. it does not restrict the generality of the
gravitational field. More specifically: Using completely arbitrary
coordinates in the most general gravitational field, it is always
possible to introduce new coordinates fulfilling the gauge
condition \ebeq{4}. In other words, Einstein's Nov 4 paper was already
generally covariant, if not manifestly so in the physical sense
that in every gravitational field one can choose coordinates
fulfilling \ebeq{4} and in these coordinates gravitation is governed by
\ebeq{2} in his Nov 4 version. The only physical reason for the
requirement of general covariance is: not to restrict the
generality of the gravitational field. {\em Manifest} general
covariance is a matter of predilection. It has the advantage of
making the non-restriction obvious i.e. manifest. With sufficient
technical skill it is always possible to find the manifestly
generally covariant formulation, the physical contents of the
theory being unchanged.

{\bf Nov 13, 1915} With a postcard, Hilbert invites Einstein to
attend his lecture at G\"ottingen scheduled for Nov 16, where he
promises to give the solution of Einstein's `great problem'. With
`great problem' Hilbert probably meant the state of affair when
Einstein was lecturing in G\"ottingen in June/July. Obviously, he
had not yet absorbed Einstein's Nov 4 page proofs, and had not
realized, Einstein himself had already solved most of his `great
problem'. (Only the discrepancy with the requirement \ebeq{3}
remained.) Hilbert has brought this postcard to the post office in
the night only, since it bears a postmark of Nov 14, 6-7 a.m.

{\bf Nov 13, 1915} After having posted the above postcard, Hilbert
writes a second postcard, which again has received a postmark of
Nov 14, 6-7 a.m. Obviously Hilbert has now studied Einstein's Nov 4
submission. Hilbert writes on this second postcard:

{\em
``As far as I understand your recent work, the solution given by you
is completely different from mine, in particular since with me
$[$the energy expression$]$ necessarily contains the electric
potential. Continued on page I with the invitation to come to here
at Tuesday 6 p.m."} With the Roman numeral `I', Hilbert refers to his
first postcard.

{\bf Nov 15, 1915} With a postcard, Einstein declines
Hilbert's invitation for the Nov 16 talk because of fatigue and stomach-ache,
and expresses his hope later to read the printed version of the talk.

{\bf Nov 16, 1915} Hilbert gives his talk at the Mathematical
Academy of G\"ottingen.

{\bf Nov 16, 1915} Hilbert
writes Einstein a postcard, which is now lost. So we
do not know the exact date, which could vary by one day,
nor exactly what was written on it. According
to Wuensch%
\ebnercite{Wuensch}
the postcard contained the following three formulae
\begin{gather*}
\tag{5a}
H = K + L\\
\tag{5b}
\delta \int (K+L)\sqrt{g}\;dw =0\\
\tag{5c}
\sqrt{g}\left( K_{\mu\nu}-{\scriptstyle\frac{1}{2}}Kg_{\mu\nu}\right)+
\frac{\partial L\sqrt{g}}{\partial g^{\mu\nu}}=0
\end{gather*}
\ebeq{5c} are the correct field equations of General Relativity,
including the trace term $-{\scriptstyle\frac{1}{2}}Kg_{\mu\nu}$.
The minus sign under the square root is absent giving the imaginary
unit $i$ which drops out of the equation.
$K_{\mu\nu}$
is the Ricci tensor, in the tradition of Gauss for German
{\em Kr\"ummung} = curvature. So $K$ is the Ricci scalar.
\ebeq{5b} is Hamilton's principle of least action. $H$ is the
Lagrangian, which by \ebeq{5a} is the sum of the Lagrangian $K$ of
the gravitational field and the Lagrangian $L$ of everything else
(matter, radiation). Since in \ebeq{5} Hilbert did not restrict
$L$ he still considers general matter (or radiation). Einstein's later reproach
concerning Hilbert's treatment of matter in a letter
to Ehrenfest%
\ebnercite{Ehrenfest}
as `unnecessarily special' and in a letter to Weyl%
\ebnercite{Weyl}
as `childish' is thus completely unjustified. However, it is true,
that later in his page proofs, Hilbert specialized $L$ to the
Lagrangian of Mie's electrodynamics, in which today nobody
believes. Hilbert denotes the coordinates by $w_a$ ($w$ = world
coordinates instead of $x^{a}$ as is usual today), so $dw$ is the
product of the four coordinate differentials in the fourfold
integral in \ebeq{5b}, and $\sqrt{g}\;dw$ is the invariant volume
element of space-time. As is well known in General Relativity,
given a Lagrangian $L$ of matter (or radiation), its
energy-momentum tensor is
\begin{equation*}
\tag{6}
T_{\mu\nu}= \frac{1}{\sqrt{g}}\frac{\partial L\sqrt{g}}{\partial g^{\mu\nu}}.
\end{equation*}
The coupling constant $\kappa$ does not occur in \ebeq{5} as it can
be put to unity while choosing suitable units.

It is curious that this Nov 16 postcard of Hilbert to Einstein is lost,
while the more trivial Nov 13 and Nov 19 postcards are preserved.

Of course it is controversial what was on the postcard.
A discussion will be found in the following section.

{\bf Nov 18, 1915} Einstein
writes to Hilbert, acknowledging Hilbert's Nov 16 postcard:

{\em
``Dear Colleague! The system, given by you, coincides -
as far I can see - exactly with what I have found in the
last weeks and have sent to the Academy. The difficulty was
not to find generally covariant equations for the $g_{\mu\nu}$;
for this is easy with the help of the Riemann tensor. But it was
difficult to recognize, that these equations are a
generalization, and indeed a simple and natural one, of
Newton's law. This I achieved only in the last weeks
(I have sent you my first communication). The
only possible generally covariant equations,
which now have turned out to be the correct ones, I
have taken into consideration with my friend Grossmann.
Only with a heavy heart we parted with them, because the
physical discussion has seemed me to have given the
incompatibility with Newton's law. - Main point, the difficulties
are now overcome. Today I send the Academy a paper, in which I deduce
quantitatively from General Relativity, without any additional ad hoc hypotheses,
Mercury's perihelion motion, as discovered by Leverrier. Thus far no
other theory of gravitation has achieved that. With best greetings
your Einstein."}

The sentence `I have sent you my first communication' clearly shows
that Einstein refers to his Nov 4 paper he has sent to Hilbert at Nov 7,
where \ebeq{2} with the Ricci tensor is used. Probably he has found
that equation several weeks ago, but has submitted it only at Nov 4. Thus
in his first reaction, Einstein only recognized in \ebeq{5c} that Hilbert has used
the Ricci tensor. Probably only in the next few days, he observed
Hilbert's additional trace term
$-{\scriptstyle\frac{1}{2}}Kg_{\mu\nu}$, and Einstein began to
experiment with additional trace terms also.

The passage `But it was
difficult to recognize, that these equations are a
generalization, and indeed a simple and natural one, of
Newton's law.' is a heavy and very justified reproach
against Hilbert. By good luck Hilbert has taken the Ricci scalar $K$
as the Lagrangian of the gravitational field and by working out
the corresponding Euler-Lagrange-equations \ebeq{5c} (which of course is
a tremendous achievement) for the first time
has found the correct field equations of General Relativity. But because
he did not show that these equations have Newton's theory as a limit,
he does not know if his equations \ebeq{5c} have possibly something
to do with gravity. Hilbert had not even a change to do so because
he specializes to Mie's electrodynamics and thus was not in a position to
check if a point mass or a fluid moves according to Newton's gravity,
as Einstein has done. It seems Hilbert did not even feel an obligation to do so.

{\bf Nov 18, 1915} Einstein
submits his paper explaining Mercury's perihelion motion.\\
The biographer Pais%
\ebnercite{Pais}
writes: `This discovery was, I believe, by far the strongest
emotional experience in Einstein's scientific life, perhaps in
all his life. Nature had spoken to him.' Einstein writes to Ehrenfest:
`For a few days I was beside myself with joyous excitement',
and to Fokker that his discovery has given him palpitations of the heart.

{\bf Nov 19, 1915} On a postcard, Hilbert congratulates Einstein
enthusiastically and in very friendly words for his explanation
of Mercury's perihelion motion.

{\bf Nov 20, 1915} Hilbert submits his manuscript which was then
printed on his page proofs%
\ebnercite{PageProofs}%
.

{\bf Nov 25, 1915}
Einstein%
\ebnercite{EinsteinNov25}
also arrives at the correct equations of General Relativity
including the trace term:
\begin{equation*}
\tag{7}
R_{ab}= -\kappa \left( T_{ab}-{\scriptstyle\frac{1}{2}}g_{ab}T\right)
\end{equation*}
at Berlin, but without citing Hilbert. To be in agreement with
Newton's theory in the limit of weak fields one must have:
\begin{equation*}
\tag{8}
\kappa = 8\pi \gamma c^{-4}
\end{equation*}
where $\gamma$ is Newton's constant of gravitation, and $c$
is the velocity of light.
\ebeq{7} is equivalent to \ebeq{5c}. Einstein does not give an
explanation how he has now found the additional trace term.
He has only checked that the conclusion $T=0$ is no longer possible.

{\bf Nov 26, 1915} Einstein
writes a letter to his friend Zangger accusing Hilbert,
without naming him explicitly, in drastic words:
{\em
``The theory has unique beauty. Only one colleague
has understood it really, but he tries in a tricky way
to `nostrify' it (an expression due to Abraham).
In my personal experience I have not learnt any better
the wretchedness of the human species as on occasion of this
theory and related to it. However, that does not
concern me in the slightest."}
and in Nov 30, 1915 he writes Besso
{\em ``Colleagues behave nastily."}

It is difficult to understand such harsh words of Einstein.
He must must have been extremely angry having worked for eight years
to the solution of his `great problem' and Hilbert in only a few weeks elegantly has
found the solution before him.
Einstein's fury shows the Hilbert's Nov 16 postcard was of
considerable help to him.

On the other hand, Hilbert writes with great admiration about Einstein, e.g.%
\ebnercite{Hilbert1924}
{\em ``the publications of Einstein, which are always rich in new
approaches and ideas."}

{\bf Dec 2, 1915} Einstein's Nov 25 paper appears. Certainly
Hilbert got a copy and very probably he was angry because
Einstein did not credit him for his Nov 16 postcard with
the correct equations \ebeq{5c}, and Hilbert has sent Einstein a letter
expressing his grievance. Such a letter, which should have
existed because of Einstein's  mentioning of a `certain resentment'
at Dec 20, 1915, is also lost.

{\bf Dec 6, 1915} Hilbert receives his page proofs from the
printery. Most probably Einstein received a copy from Hilbert in
the following day.

{\bf Dec 20, 1915} Einstein asks Hilbert for reconciliation:
{\em ``There has been a certain resentment between us, the cause of which
I do not want to analyze any further. I have fought against the
feeling of bitterness associated with it, and with complete success.
I again think of you with undiminished  kindness and I ask you to
attempt the same with me. It is objectively a pity if two real
guys that have somewhat liberated themselves
from this shabby world are not giving pleasure to each other."}\\
In view of their friendly correspondence before, that offer for reconciliation
is  astonishing. Obviously other letters or postcards have been lost.
Most probably Hilbert was angry Einstein did not reference him in the Nov 25
submission. However, Hilbert could not have uttered such a reproach outright,
because then Einstein had nothing to analyze. Sommer%
\ebnercite{Sommer}
suspects Einstein's offer was not an act of pure love. At that time Hilbert was
much more famous than Einstein. A published reproach by Hilbert would
have been disastrous for Einstein.

{\bf Mar 31, 1916} The printed version%
\ebnercite{HilbertPublished}
of Hilberts page proofs appear. He has made several revisions,
retaining his date of submission, Nov 20, 1915, nor adding a date of revision.

{\bf March 7, 1918} Hilbert sends Klein his page proofs asking him
to be careful to them sending them back because he had no other
records.

{\bf 1924} Hilbert's republication%
\ebnercite{Hilbert1924}%
, claiming `rearrangements and editorial changes' only,
but in fact abandoning his own derivation and switching to one already
standard at that time, using Riemannian normal coordinates.

{\bf 1943} Hilbert's death.

{\bf 1943-1967} Hilbert's documents are deposited in a adjoining room
of the office of the Mathematical Institute at G\"ottingen%
\ebnercite{Wuensch}%
.

{\bf 1966} About 130 letters written by Einstein, Planck, Born, Nernst, Debye,
Sommerfeld, Weyl, Courant, Ehrenfest addressed to Hilbert have been sent to
New York in order to copy them. They have been sent back but not deposited
in the office again. Finally in the year 2000 they were found by Sommer%
\ebnercite{Sommer}
on an attic of the widow of an assistant of Courant.

{\bf 1967} The remaining documents (not having been sent to New York) came
to the department of rare documents of the library of the university at G\"ottingen
(called the `G\"ottingen archive'%
\ebnercite{OrderFromArchive}
in the following).

{\bf 1985} The correspondence between Hilbert and Klein is published%
\ebnercite{Kleinschwierig}%
.

{\bf 1994} Corry detects Hilbert's page proofs at G\"ottingen. He
orders copies which are sent to the Max Planck institute for the
History of Science at Berlin where he studies them in the
following year.

{\bf Nov 14, 1997} {\em Science} article by Corry, Renn and Stachel%
\ebnercite{Science}%
, without mentioning the cut-off.

{\bf 1999} Tilman Sauer%
\ebnercite{Sauer1}%
for the first time mentions the cut-off. Also
Renn and Stachel, but only in a footnote%
\ebnercite{Sommer}%
.

{\bf 2003} F. Winterberg%
\ebnercite{Winterberg}
for the first time publishes the suspicion the
cut-off was the deed of a forger. Also in 2004 by
C. J. Bjerknes%
\ebnercite{Bjerknes}%
.

{\bf Mar 2005} Wuensch's book%
\ebnercite{Wuensch} appears giving strong evidence the cut-off has been
done in recent years, most probably by a skilled historian,
because curious traces can be seen on Hilbert's page proofs, which
find a plausible explanation only by attempts of cover-ups,
presupposing knowledge of detailed historical facts.
\section{What was on Hilbert's Nov 16, 1915 postcard to Einstein?}
It is curious this postcard is lost, while his more trivial
Nov 13 and Nov 19 postcards are preserved. However, we can safely infer
what was essentially written on Hilbert's Nov 16 postcard, namely
from Einstein's reactions:

Einstein Nov 18:\\
{\em
``The system, given by you, coincides -
as far I can see - exactly with what I have found in the
last weeks and have sent to the Academy. The difficulty was
not to find generally covariant equations for the $g_{\mu\nu}$;
for this is easy with the help of the Riemann tensor. But it was
difficult to recognize, that these equations are a
generalization, and indeed a simple and natural one, of
Newton's law. This I achieved only in the last weeks."
}

This clearly demonstrates Hilbert has sent explicit field equations
for general relativity (and e.g. not only a Lagrangian density).

The passage:
{\em
``The system, given by you, coincides -
as far I can see - exactly with what I have found in the
last weeks and have sent to the Academy."
} opens the possibility Hilbert has sent incorrect field equations such as
\ebeq{2} lacking the trace term. However, this can be excluded by Einstein's
angry reaction in his letter Nov 26 to his friend Zangger. Hilbert had had no
chance to `nostrify', if he had sent incorrect field equations.

Rather, in his first reaction, Einstein did not observe Hilbert's additional
trace term but only Hilbert's use of the Ricci-tensor, so Einstein erroneously
believed Hilbert's system coincided with his own.

Some historians believe Hilbert, after having seen the correct equations \ebeq{7} in
Einstein's paper, he had {\em copied} from Einstein when he has added the correct field
equations for his final printed version somewhen after Dec 6. However, this can safely
be excluded because Einstein did not give any method how he finally arrived at the
correct equations \ebeq{7}, and in particular he did not derive them from a Lagrangian
density. So it would be an extreme risk for Hilbert to {\em copy} from Einstein,
since he could
not know if Einstein's equations coincide with those following
via variational calculus from his Lagrangian density.

\section{How Einstein has found his equations}
Einstein was dissatisfied with his equations \ebeq{2}, because they led
to the unphysical conclusion $T=0$. It is curious Einstein did not
attempt a trace term on the left or right hand side of \ebeq{2}. It am sure,
sooner or later, he alone or perhaps by a hint by someone less
famous than Hilbert, had he attempted such trace terms. Perhaps, Einstein's
Nov 26, 1915 letter to Zangger can be explained by his anger he has not come
to this simple idea by his own.

\subsection{\bf Was Grossmann a poor mathematician?}
Why that hint did not come from Einstein's friend
Grossmann, who should have known the contracted Bianchi identities
\begin{equation}
\tag{9}
(R^b_a-{\scriptstyle \frac{1}{2}}Rg^b_a)_{;b}=0
\end{equation}
which follow from the Bianchi identities,
discovered by Bianchi%
\ebnercite{Bianchi}
in 1902.

In 1899 a German translation (first edition) of Bianchi's `Differentialgeometrie'
appears, of course without the Bianchi identities. In his second 1910
German edition, Bianchi has omitted differential geometry for curved
space with $n>2$. Obviously such a subject seemed too devoid of
any application to be included in a textbook. Only in a footnote,
Bianchi has added the symmetry properties of the Riemann tensor, which are
trivial for $n=2$. His Bianchi-identities seemed too uninteresting
to be included even in that footnote.

For the years 1850-1915, I have scanned all German
journals, the Sommerville bibliography%
\ebnercite{Sommerville}%
, and all German and non-German
text books reviewed in Acta Mathematica, and I did not
find any mentioning of Bianchi's (contracted) identities.
1869 was the start of the excellent reviewing journal
``Jahrbuch \"uber Fortschritte der Mathematik" reviewing most
German and non-German mathematical works. Bianchi's discovery
was reviewed in Vol 33(1902) but completely un-conspicuous
among the enormous amount which was published in mathematics
already at that times. Until 1915 not a single mentioning
of the identities. They appear for the first time
in 1918 in Schouten's textbook%
\ebnercite{Schouten1}%
.

Also Hilbert had no chance to know \ebeq{9}. So the assumption that he had used
it for his derivation of the explicit field equations are untenable.
However, {\em after} he had derived the field equations, because of Theorem
III of his page proofs, he new the contracted Bianchi identities implicitly.
But definitely, they were not among his explicitly known theorems.

\subsection{\bf Has Einstein found his equations independently from Hilbert?}

In Nov 11, 1915, Einstein still had the equations in the form \ebeq{2},
lacking the trace term. He was
dissatisfied because, using the gauge \ebeq{4}, which is always possible, he could
conclude
\begin{equation*}
\tag{10}
T = \text{ const.}
\end{equation*}
i.e. $T\equiv 0$ by integrating to a vacuum point.

In his Nov 25 paper, Einstein does not give a method how he has
arrived at the additional trace term in \ebeq{7}. He merely shows that with the trace term,
conclusion \ebeq{10} can no longer be drawn. He does not show \ebeq{3}, which amounts to
proving the contracted Bianchi identities \ebeq{9}. Thus in Nov 25, Einstein
did not know his theory was already complete, though he%
\ebnercite{EinsteinNov25}
writes on p. 847:
{\em ``Thus finally, general relativity is a closed logical edifice.
The relativity postulate in its most general formulation, according to
which coordinates are irrelevant parameters, leads with absolute necessity
to a uniquely specified theory of gravitation"}.
Obviously, he relied on Hilbert's
Nov 16 postcard for this (partially incorrect) conviction.
And this again excludes the possibility
Hilbert has sent with his Nov 16, 1915,
postcard only undetermined coefficients in front of
the Ricci tensor and in the trace term.

\section{A short introduction to the theory of invariants}
Hilbert's derivation of the Einstein equations are essentially
based on invariant theoretical arguments.

\subsection{\bf Algebraic invariants}
The invention of analytical geometry, i.e. the application of
algebra and infinitesimal calculus to geometry, was a major
progress in solving geometrical problems. However, it had a
great disadvantage, because it required the introduction of a
coordinate system which is arbitrary and not or not uniquely related
to the geometrical objects under consideration. This was the birth
of the theory of invariants. We mention the names of Sylvester
and Cayley. It was mainly an algebraic invariant theory and only a binary one,
applicable to the Euclidean $n=2$ dimensional plane. The best known algebraic
invariant is the sum of the squares of the coordinate difference of
two points being the geometrical objects, an invariant which has the geometrical
interpretation as the square of the distance between these points.
Algebraic invariant theory culminated with the textbooks 1872 by Clebsch and
1885 by Gordan%
\ebnercite{Clebsch}%
.
Most mathematicians at that time and in
particular Hilbert contributed to that field.
Hilbert has become famous for the first time  by his finiteness theorem
of the system of algebraic invariants
(solving Gordan's problem for $n>2)$.

\subsection{\bf Gauss's measure of intrinsic curvature}

The first milestone in {\em differential} invariants
was 1828 Gauss: {\em Disquisitiones generales circa superficies curvas}%
\ebnercite{Gauss}%
, where
he considered a general $n=2$ dimensional curved surface embedded in
3-dimensional Euclidean space. (Disregarding global topology this is the
most general 2-dimensional Riemannian space.) He introduced the following differential
invariant:
\begin{gather*}
\tag{11}
{\scriptstyle
k =
\frac{1}{4}(EG-F^2)^{-2}
\Big[
E\big(
\frac{\partial E}{\partial q}\frac{\partial G}{\partial q}
-2 \frac{\partial F}{\partial p}\frac{\partial G}{\partial q}
+( \frac{\partial G}{\partial p} )^2
\big)+
F
\big(
\frac{\partial E}{\partial p}\frac{\partial G}{\partial q}
-\frac{\partial E}{\partial q}\frac{\partial G}{\partial p}
-2\frac{\partial E}{\partial q}\frac{\partial F}{\partial q}
+4\frac{\partial F}{\partial p}\frac{\partial F}{\partial q}
-2\frac{\partial F}{\partial p}\frac{\partial G}{\partial p}
\big)+
} 
\\
{\scriptstyle
+G
\big(
\frac{\partial E}{\partial p}\frac{\partial G}{\partial p}
-2\frac{\partial E}{\partial p}\frac{\partial F}{\partial q}
+( \frac{\partial E}{\partial q} )^2
)
-2(EG-F^2)
(
\frac{\partial^2E}{\partial q^2}
-2\frac{\partial^2F}{\partial p\partial q}
+\frac{\partial^2G}{\partial p^2}
\big)
\Big]
} 
\end{gather*}
where $E=g_{11}, F=g_{12}, G=g_{22}$ and $p=x^1, q=x^2$
are arbitrary curvilinear coordinates on the surface. The
invariant has the geometrical significance of $k$ what is now
called the `Gaussian measure of curvature'. Taking a finite piece
of the surface, he collected all unit normal vectors
(orthogonal to the surface) and fixed them at the origin of
a unit sphere. The area of their end points he called
total curvature ({\em curvatura integra}) of the surface.
$k$ (at a point $P$) is the quotient of that area by the area
of the surface itself, taking an infinitesimal neighbourhood of $P$.
$k$ is also
the product of the minimal and maximum exterior curvature
of geodesic lines emanating from $P$. He showed that $k$
is an invariant against bending (think of a thin material surface
resisting to stretch and tear) and of course against rotation
and translation. Since every $n=2$ Riemannian space can be
(locally) embedded in $R^3$, `bending' (which is not necessarily
a continuous
process, and is also referred to as `developing one surface unto another'
e.g. developing a cylinder unto a plane) is synonymous with
`isometry'. Thus \ebeq{11} is a `bending invariant', and $k$ was called
`internal curvature of the surface'. Trivially, since no assumptions
about the choice of the coordinates $p,q$ have been made, \ebeq{11} is
also an invariant against arbitrary coordinate transformations.
Gauss called this invariance properties {\em theorema egregium}
(remarkable theorem.)

\subsection{\bf Riemann and Christoffel: the $n$-dimensional case}

In 1854 in his {\em Habilitationschrift}%
\ebnercite{Riemann}
(where because of the audience he
could use words only) Riemann introduced qualitatively the concept of
an $n$ dimensional Riemannian space, by considering the quadratic form
\begin{equation*}
\tag{12}
ds^2=\sum_{a,b=1}^n g_{ab}dx^a dx^b
\end{equation*}
describing the metric.
He generalizes Gauss's work by constructing
all geodesic
(2-dimensional) surfaces emanating (with all orientations) from a point $P$,
considering their $k$ given by \ebeq{11}.

The hard formula work was done in 1869 by E.B. Christoffel%
\ebnercite{Christoffel}
where he introduced his 3-index-symbols, the 4-index-symbols (now
called Riemann tensor), the covariant derivative,
permitting to construct invariants and covariants, i.e. tensors, of
arbitrary order.
(Only later and posthumously, it was known that Riemann had done
a substantial part of this work before%
\ebnercite{ParisPrize}%
.)
Gauss's $k$ is regained as the (only independent) component of the Riemann tensor:
\begin{equation*}
\tag{13}
k = g^{-1}R_{1212}
\end{equation*}

\subsection{\bf Gaussian invariants}
A Gaussian invariant of order $G$ of $n$ variables $x^a$ is a function
$J(x^a, g_{ab}, \cdots, g_{abk_1\cdots k_G}$) depending on arbitrary
functions $g_{ab}(x^k)$ and their partial derivatives (denoted by
additional indices) up to the order $G$, with the condition that the
value of $J$ remains unchanged if we introduce new variables $x^{\prime\mu}$
(coordinate transformation) whereby the $g'_{\mu\nu}$ are
found by the well known formulae for transforming a metrical tensor, and their
derivatives are now calculated with respect to the new coordinates.
Symbolically we could write
\begin{equation*}
\tag{14}
J(x^a,{\cal G})=J(x'^a,{\cal G}')
\end{equation*}
where $\cal G$ summarizes all metrical variables (including their derivatives).
On both sides $J$ is the same mathematical function (`form invariance')
and the primed und unprimed arguments on both sides are connected by
the coordinate transformation, and are taken at the same objective geometrical point,
having two different coordinates ($x^a$ and $x'^a$).\\

The Ricci scalar is a Gaussian invariant of order $G=2$. In this case the function
$J$ is the (form-invariant, i.e. coordinate independent)
prescription how the Ricci scalar is calculated.
There are no Gaussian invariants of lower ($G<2$) order.

As an exercise we will prove, that an invariant $J$ cannot depend explicitly on
the coordinates $x^a$.
Taking a fixed invariant $J$, \ebeq{14} is an infinite set of
numerical equalities, where the set runs over all possible choices for the
functions $g_{ab}$, all values of the coordinates $x^a$ and all possible
coordinate transformations. Arrange the elements of this set in pairs,
where two members of the pair differ by the choice of the coordinate
transformation only, namely
by an additional translation ($x''^a=x'^a+c$) with a fixed
chosen $c$ and $a$. Both members have the same left hand side of \ebeq{14}. By writing
down the transformation law for a metric tensor for $'$ and for $''$ (and
for its derivatives) one sees that for the collection of numerical values
(denoted by ${\cal G}$) there holds ${\cal G}'={\cal G}''$.
Thus the right hand side of the members differ
only in the variable $x'^a$ by $c$. Since $c$ is arbitrary, $J$ cannot depend on $x^a$.

\subsection{\bf Beltrami invariants}
To a quantity constructed from the components of the metrical tensor
(and its partial derivatives) alone
(also called a `concomitant' of the metric) we give the specifier `Gaussian'.

To subsume Gaussian tensors (also called Gaussian `covariants') to the concept
of invariants, we have to consider Beltrami invariants:

A Beltrami invariant (in the literature also denoted by the strange name
`differential parameter of the first kind') of orders $(G, B)$ is a function $J$ which
additionally depends on a scalar function $\varphi(x^a)$ and its partial derivatives
$\varphi_a, \cdots \varphi_{k_1\cdots k_B}$ up to the order $B$.\\
(Note that a scalar function is also transformed, i.e. is
replaced by a new mathematical function $\varphi^\prime$:
\begin{equation*}
\tag{15}
\varphi(x^a)=\varphi^\prime(x^{\prime a})
\end{equation*}
though it is usual in physics
to omit the prime on $\varphi$ distinguishing different functions.)\\
Beltrami (based on Lam\'{e}) has found the square of the gradient
($g^{ab}\varphi_a\varphi_b$) and the Laplacian ($g^{ab}\varphi_{a;b}$),
what is now called the first ($G=0, B=1$) and second ($G=1, B=2$) Beltrami invariant.

The concept is enlarged to include $S$ scalar functions $\varphi^s$.
A Gaussian {\em covariant} (e.g. the metrical, Riemann or Ricci tensor) can be
viewed as a Beltrami invariant  which is contracted with the gradients (B=1)
of $S$ ($S$ = order of the tensor) scalar fields. Since we have to deal
with symmetrical second rank tensors $T_{ab}$ only, we can restrict ourselves to
$S=B=1$, since the invariant $T^{ab}\varphi_a\varphi_b$ for arbitrary
$\varphi_a$ determines the tensor $T_{ab}$ uniquely.

We say that there are $N=N(n,G,B,S)$ Beltrami invariants $J_1,\cdots,J_N$
if every Beltrami invariant $J$  of this or lower order can be expressed as a
function $\cal J$ of them:
\begin{equation*}
\tag{16}
J={\cal J}(J_1,\cdots,J_N)
\end{equation*}
and these $N$ invariants are independent, i.e. no one
can be expressed (in the sense of \ebeq{16}) by the remaining $N-1$ ones.

The $S$ scalar functions (e.g. take $J=J(\varphi)=\varphi$)
are themselves invariants (though not differential invariants).
It is usual to discard them as trivial.
When the number $N$ of
a complete system of independent invariants $J_1,\cdots,J_N$ is calculated,
they are not counted. Even more radically, an invariant is called a
Beltrami-invariant only if it does not depend on the $\varphi^s$ (but
on their derivatives $\varphi^s_{a_1\cdots a_B}$ only).

This terminology has the consequence that a Beltrami invariant $B=0$ is
a Gaussian invariant.

As an exercise, we prove this is a good practice.
For clarity we distinguish free Beltrami invariants
(`f-invariants') from general Beltrami invariants (`g-invariants')
which may depend on the $\varphi^s$. We will prove the
following lemma: When $J_1,\cdots,J_N$
is a complete system of independent f-invariants,
then an arbitrary g-invariant is again given by a function $\cal J$ in \ebeq{16}
with the $\varphi^s$ as additional arguments.

For simplicity we take the typical case the g-invariant $J$ fulfills
the equations:
\begin{equation*}
\tag{17}
J({\cal G},\varphi,\varphi_a) = J({\cal G}',\varphi',\varphi'_a)
\end{equation*}
i.e. $S=1, B=1$
As in our previous exercise, \ebeq{17} is an infinite set of numerical
equalities. Now we show the same set also corresponds to
the equations of the following f-invariant (choosing a fixed $c$):
\begin{equation*}
\tag{18}
\bar{J}({\cal G},\varphi_a;c):= J({\cal G},c,\varphi_a) = J({\cal G}',c,\varphi'_a)
\end{equation*}
Consider one element of the set, specified by coordinates $x^a$ and a function $\varphi$
(with corresponding values $x'^a$ and $\varphi'(x'^a)=\varphi(x^a)=:C$ on the right hand side
of \ebeq{18}). When $c=C$ \ebeq{18} is fulfilled. Otherwise pick the element with function $\Phi=\varphi+c-D$
out of the set (same $x^a$). $\Phi$ and $\varphi$ have the same derivatives. So the picked
element says \ebeq{18} is again fulfilled.\\
Since $c$ is arbitrary, $\bar{J}$ are an infinite set of f-invariants
(except when $\bar{J}$ does not depend on $c$, i.e. $J$ was already an
f-invariant). Since sum, difference, constant multiples of invariants
are again invariants, derivatives (to arbitrary order) of $\bar{J}$
with respect to $c$ are again f-invariants. Assuming power series
for $J$ in its middle argument, we have proved our lemma.

\section{How Hilbert has found the Einstein equations}

\subsection{\bf Hilbert's  invariant theoretical prerequisites to find the Einstein equations}
Hilbert derives the Einstein equation by a variational principle with
a Lagrangian density
\begin{equation*}
\tag{19}
L=R+L_{\text{em}}
\end{equation*}
(Hilbert writes $H=K+L, \quad L_{\text{em}}$ is an electromagnetic Lagrangian density,
later specialized to that one of Mie's theory.)
Thus Hilbert has to calculate the variational derivative
\begin{gather*}
\tag{20}
\frac{\delta\sqrt{g}R}{\delta g^{ab}}=\frac{\partial \sqrt{g}R}{\partial g^{ab}}-
\sum_{k=1}^n \frac{\partial}{\partial x^k}\; \frac{\partial \sqrt{g}R}{\partial g_k^{ab}}+
\sum_{k,l=1}^n \frac{\partial^2}{\partial x^k \partial x^l}\;
   \frac{\partial \sqrt{g}R}{\partial g_{kl}^{ab}}=:\sqrt{g}G_{ab}
\end{gather*}
where for reasons of convenience we have introduced the quantities $G_{ab}$
for the result of this calculation. To derive the Einstein equations, he
had to show:
\begin{equation*}
\tag{21}
G_{ab}=R_{ab}-{\scriptstyle\frac{1}{2}}R g_{ab}.
\end{equation*}

In Theorem III of his page proofs he shows that $G_{ab}$ is a second
rank tensor.
Obviously it is a symmetric one.

$R$ contains the second derivatives $g^{\mu\nu}_{kl}$ only
linearly and the first derivatives $g^{\mu\nu}_k$ only bilinearly
with coefficients containing the zero-th derivatives only. Thus inspection
into the variational procedure \ebeq{20}
easily leads to the conclusion $G_{ab}$ can contain
the second derivatives only linearly.
(Note that $g^{ab}$ are rational functions of the $g_{ab}$ and vice versa,
with $g$ or $1/g$ in the denominator.)

In invariant theoretical language, Hilbert's assumption \ebeq{1}
says that up to order $G=2$, $B=1$,
$S=1$ there are only 3 independent invariants. As these we can take:
\begin{equation*}
\tag{22}
J_1=R,\quad J_2=g^{ab}\varphi_a \varphi_b, \quad J_3=R^{ab}\varphi_a\varphi_b
\end{equation*}
which are independent.
So every invariant $J$ of that type is a
function of them: $J={\cal J}(J_1, J_2, J_3)$. The invariants \ebeq{22} are
linear in the second derivatives of the metric. If $J$ should also have this
property and be bilinear in the $\varphi_a$, it follows (assuming
power series and because the $J_i$ can assume independently any values) that
\begin{equation*}
\tag{23}
G^{ab}\varphi_a\varphi_b=J=[{\ebnerboldmath \alpha} R^{ab}+
{\ebnerboldmath \beta} R g^{ab}+{\ebnerboldmath \lambda} g^{ab}]\varphi_a\varphi_b.
\end{equation*}
Since the $\varphi_a$ are arbitrary, and since both $G^{ab}$ and $[\quad]$ in \ebeq{23} are symmetric
we have:
\begin{equation*}
\tag{24}
G_{ab}={\ebnerboldmath \alpha} R_{ab}+{\ebnerboldmath\beta} R g_{ab}+{\ebnerboldmath\lambda} g_{ab}
\end{equation*}

Note that the function $\cal J$ (for any chosen $J$) in \ebeq{16} does not depend
on any of the arguments of an invariant, i.e. not on $g^{ab}, g^{ab}_c \cdots$.
This fact, trivial for Hilbert, was perhaps the essential stumbling-block,
why Hilbert's derivation of the Einstein equations was not understood thus
far since 90 years.

For a fixed $n$ (dimension of space)
$J$ is a fixed invariant. Thus
${\ebnerboldmath \alpha, \ebnerboldmath\beta, \ebnerboldmath\lambda}$
are numerical constants, possibly depending on $n$ only.

Thus Hilbert's method how he has found the Einstein equations `easily without calculation'
becomes obvious: Take a metric $g_{ab}$ as simple as possible, but non-trivial enough so
that the constants
${\ebnerboldmath\alpha, \ebnerboldmath\beta, \ebnerboldmath\lambda}$
are determined uniquely
while \ebeq{20} is evaluated.
Otherwise the metric is completely arbitrary. It had not to be physically
meaningful in any way. That is done in the following subsection.

\subsection{\bf How Hilbert has found the {\em explicit form} of the Einstein equations}
From the G\"ottingen archive%
\ebnercite{OrderFromArchive}
one can obtain a Xerox copy of
a microfilm of Hilbert's private folder ``{\em Zur Elektrodynamik}".
(Hilbert erroneously%
\ebnercite{HilbertErroneously}
believed to have shown the
electromagnetic phenomena to be a consequence of gravitation.)
In this folder Hilbert's derivation can be found on the page with Archive-number 32.
Here he collects ``{\em Formeln}" he has derived for the Ricci-tensor and
the Ricci-scalar in the case of a general diagonal metric:
\begin{gather*}
\tag{25}
R_{ii}= {\scriptstyle \frac{1}{2}}\sum_{k,(k\neq i)}
\Big\{
g^{kk}(g_{kkii}+g_{iikk})-\\
{\scriptstyle\frac{1}{2}}g^{kk}
\left( 2g^{ii}g_{iii}g_{kki}+g^{kk}g_{kki}g_{kki}+2g^{kk}g_{iik}g_{kkk}+g^{ii}g_{iik}g_{iik}\right)
+{\scriptstyle\frac{1}{2}}g^{kk}\sum_\rho g^{\rho\rho}g_{ii\rho}g_{kk\rho}
\Big\}.
\end{gather*}

Then he mentions a ``{\em Spezialfall}" (special case)
\begin{equation*}
\tag{26}
g_{11}=g_{22}=g_{33}=1, \quad g_{44}=\gamma,\quad g^{44}=1/\gamma
\end{equation*}
and for this special case he gives the Ricci scalar
\begin{equation*}
\tag{27}
R = \sum_{i<4}(\gamma^{-1}\gamma_{ii}-{\scriptstyle\frac{1}{2}}\gamma^{-2}\gamma_i^2).
\end{equation*}
We simplify Hilbert's special case even a bit more by assuming
\begin{equation*}
\tag{28}
\gamma = \gamma(x) \text{ with }x\equiv x^1.
\end{equation*}
Introducing $\kappa = g^{44} = 1/\gamma$ (or starting afresh from the
definitions in textbooks; Hilbert's formulae differ
by a minus sign from the convention in Landau-Lifshitz%
\ebnercite{LandauLifshitz}%
) one immediately finds ($' =$ derivative with respect to $x$):
\begin{gather*}
\tag{29}
\sqrt{g}=\kappa^{-\frac{1}{2}}\\
\tag{30}
R = -\kappa^{-1}\kappa''+{\scriptstyle\frac{3}{2}}\kappa^{-2}\kappa'^2\\
R_{44}=-{\scriptstyle\frac{1}{2}}\kappa^{-2}\kappa''+
{\scriptstyle\frac{3}{4}}\kappa^{-3}\kappa'^2
\tag{31}
\end{gather*}
Now we do the variational derivative in \ebeq{20} for $\mu=\nu=4$, which in this case yields zero:
\begin{gather*}
\tag{32}
\frac{\partial\sqrt{g}R}{\partial\kappa}
-\frac{\partial}{\partial x}\frac{\partial \sqrt{g}R}{\partial \kappa^{'}}
+\frac{\partial^2}{\partial x^2}\frac{\partial \sqrt{g}R}{\partial\kappa^{''}}
=0\stackrel{!}{=}
\\
\stackrel{!}{=}
\kappa^{-\frac{1}{2}}
\big[
 {\ebnerboldmath\alpha}
 (-{\scriptstyle\frac{1}{2}}\kappa^{-2}\kappa''
  +{\scriptstyle\frac{3}{4}}\kappa^{-3}\kappa'^2)+
 {\ebnerboldmath\beta}\kappa^{-1}
  (-\kappa^{-1}\kappa''+{\scriptstyle\frac{3}{2}}\kappa^{-2}\kappa'^2)
+{\ebnerboldmath\lambda}\kappa^{-1}
\big]
\end{gather*}
${\ebnerboldmath \alpha, \ebnerboldmath\beta, \ebnerboldmath\gamma}$
are independent of the function $\kappa$, so we
can equate coefficients of $\kappa^{\prime\prime}$ and $\kappa^\prime$
which leads to $\lambda=0$ and
\begin{equation*}
\tag{33}
\alpha+2\beta=0
\end{equation*}
(Note that we cannot check the other components $\mu\nu$ since our simple
formulae \ebeq{27} are not valid for variations of the other components of $g^{\mu\nu}$.)\\
By this simple method, Hilbert has derived the Einstein equations and has
communicated them to Einstein on his 1915 Nov 16 postcard. \ebeq{33} is equivalent to
what the contracted Bianchi identities had given, i.e. the correct trace term. For Einstein
the absolute values of $\alpha$ and $\beta$ are irrelevant, since Einstein determined the
constant of gravitation $\kappa$ by comparison with Newton's theory. Since Hilbert,
in contrast to Einstein, also gives an explicit Lagrangian for the sources of
the gravitational field, he needs the absolute values $\alpha=1$ and
$\beta=-{\scriptstyle \frac{1}{2}}$. He might have found them by calculating another
special case. We have no hint to that. Therefore, in the next subsection, we propose
another even more elegant route Hilbert might have guessed these correct values.

The derivation, just given, cannot be found in Hilbert's private folder.
Hilbert sometimes has calculated on letter sheets he has obtained from Swiss hotels
(bearing a logo of the hotel in the letter head). Paper was extremely expensive. Hilbert had
in his garden a big blackboard under a roof where he made his intermediate
calculations%
\ebnercite{ConstanceReid}%
.
Thus on his sheet 32 we
only find important formulae such as \ebeq{25}, and the indication of the `Special case'
he had used, but not the trivial calculation itself.

This derivation by Hilbert is not only historically the first one, it is also
a nice exercise for an elementary class in General Relativity, presupposing
that the field equations have the form \ebeq{24}, and fitting the constant of
gravitation by Newton's limit.

\subsection{\bf `easily without calculation'}
In this subsection we give a route how Hilbert immediately might have arrived at
\ebeq{39} using his tremendous power of
mathematical intuition. ($\lambda=0$ was obvious to him, by inspection into the
procedure \ebeq{20} of variational derivation.)

In an attempt to calculate the left most side of \ebeq{20} he writes down
\begin{equation*}
\tag{34}
\delta \sqrt{g}=-{\scriptstyle\frac{1}{2}}\sqrt{g}g_{ab}\;\delta g^{ab}.
\end{equation*}
The theory of determinants, which are the most important invariants,
was at the focus of interest at that time. Riemannian spaces for $n>2$
have been studied almost exclusively for the isotropic case
\begin{equation*}
\tag{35}
R_{ab}=\frac{R}{n}\;g_{ab},
\end{equation*}
for which F.Schur's theorem%
\ebnercite{Schur}
that $R$ is spatially constant, was famous. For variation of the metric
\begin{equation*}
\tag{36}
\delta g^{ab}=g^{ab}\delta\chi
\end{equation*}
with a spatially constant $\delta\chi$, we immediately find
\begin{equation*}
\tag{37}
\delta R = R\;\delta\chi
\end{equation*}
since $R$ scales with $g^{ab}$. So from \ebeq{20} \ebeq{24},
multiplied by \ebeq{36}, we find
\begin{equation*}
\tag{38}
1=\alpha+(\beta+{\scriptstyle\frac{1}{2}})n.
\end{equation*}
Together with \ebeq{33} we find
\begin{equation*}
\tag{39}
\alpha=1, \quad \beta=-{\scriptstyle\frac{1}{2}}.
\end{equation*}

Since in the variational procedure \ebeq{20} and in its result \ebeq{24} the
dimension $n$ of space enters only as the boundary of a formal sum, perhaps
Hilbert, using his intuition, knew that
$\alpha$ and $\beta$ cannot
depend on $n$. So he has arrived immediately from \ebeq{38} to \ebeq{39} even
without \ebeq{20}, indeed, `easily without calculation'.\\

{\footnotesize
For the reader uneasy with \ebeq{37} we proceed more formally: Start from the
middle expression in \ebeq{20}, and \ebeq{24} with $\lambda=0$. Multiply by
\ebeq{36} and integrate over the whole space of \ebeq{35}, taken as the
$n$-dimensional sphere S$^n$. Reverse the steps of partial integration,
which had led to the middle expression of \ebeq{20}. No boundary terms
appear, since we integrate over a closed space, even for a constant $\delta\chi$.
Instead, the complete differential $\delta\sqrt{g}R$ appears. Consider the
range of metrics $\chi g^{ab}$, where $g^{ab}$ is fixed and given by \ebeq{35}, and
$\chi$ is spatially constant. Thus the complete differential $\delta R$ is given
by \ebeq{37}. Push the constants $R$ and $\delta\chi$ in front of the integrals
to arrive at \ebeq{38}.
}
\section{Discussion of Hilbert's derivation}

\subsection{\bf Hilbert challenged by Zorawski and Haskins}
Sophus Lie (1842-1899) made a tremendous contribution to the theory of invariants
by recognizing that a (transformation) group (its 1-component) is uniquely
determined by
its infinitesimal transformations. Thus an invariant is already
identified when it is invariant with respect to infinitesimal
transformations. This opened the possibility to determine
the number $N$ of (independent) invariants by counting the number of
equations and the number of freedoms of the invariants, and using lemmas
guaranteeing the functional independence of the invariants thus found.
This method only gives the {\em number} of (independent) invariants,
not their (global) explicit form, except in cases where this is already known
otherwise. These numbers have been found in 1891
by Zorawski%
\ebnercite{Zorawski1}
for $n=2$ and for general $n$ in 1904 by Haskins%
\ebnercite{Haskins}%
.

Table 1 summarizes some of their results.%

\begin{table*}
\begin{tabular}{|c||c|c|c|c|c|}\hline
$N$   & $G=0$ & $G=1$ & $G=2$ & $G=3$ & $G \ge 4$\\ \hline\hline
$B=0$ &  0 & 0 &
  ${\scriptstyle\frac{1}{12}}(n-2)(n-1)n(n+3)$ & $n\frac{(n+2)!}{4!(n-2)!}$ &
                                              $n\frac{G-1}{2}\frac{(n+G-1)!}{(n-2)!(G+1)!}$ \\
& & & ${\scriptstyle N=1\text{ for } n=2}$ & ${\scriptstyle N=1\text{ for } n=2}$ & \\ \hline
$B=1$ &   1   &   0   & $n-1$  & 0 & 0\\
& & & ${\scriptstyle N=0 \text{ for } n=2}$ & ${\scriptstyle N=1 \text{ for } n=2}$ & \\ \hline
$B=2$ & 0 & $2n-1$ & $(n-1)(n-2)/2$ & 0 & 0 \\
&  & ${\scriptstyle N=3\text{ for } n=2}$  & & &\\\hline
$B=3$ & 0 & 0 & $\frac{(n+2)!}{(n-1)!3!}$ & 0 & 0 \\
\hline
\end{tabular}
\caption{Number $N$ of independent differential invariants in $n$-dimensional space ($n \ge 2$).
$G=2$ means the invariant involves up to {\em second} partial derivatives of the metrical
tensor with respect to the coordinates. $B=0$ are the Gaussian invariants, which do
not involve scalar fields. $B=1$ are Beltrami-invariants involving the {\em first}
derivatives $\varphi_a$ of a single scalar field $\varphi$. The whole table counts only
Beltrami invariants involving (derivatives of) a {\em single} scalar field ($S=1$).
A box (e.g. $G=3, B=2$) of the table does not count invariants which are dependent
(can be constructed via \ebeq{16}) by invariants in lower boxes ($G \le 3, B \le 2$).
The case $n=2$ is exceptional, so sometimes its $N$ is given separately.
}
\end{table*}

As an illustration we discuss some of the boxes. The scalar field $\varphi$, being
itself an invariant and belonging to the box $(G=0, B=0)$ is a trivial invariant and is not
counted as explained in subsection \ebeq{5.5}.

The first Beltrami invariant $g^{ab}\varphi_a\varphi_b$ is the one invariant in
the box $(G=0, B=1)$.

In the box $(G=2, B=0)$ we find the Ricci scalar $R$, which only for $n=2$ is the
only invariant of that type. Already for $n=3$ we have
3 independent Gaussian invariants. They have been given explicitly in 1873 by
Souvaroff%
\ebnercite{Souvaroff}%
:
\begin{gather*}
\tag{40}
S_1=R,\quad S_2=g^{-1}R_{kl}\varepsilon^{la\alpha}\varepsilon^{kb\beta}R_{a\alpha b\beta},\\
S_3= g^{-2}R_{klmn}
\varepsilon^{ka\alpha}\varepsilon^{lb\beta}\varepsilon^{mc\gamma}\varepsilon^{nd\delta}
R_{a\alpha c\gamma}R_{b\beta d\delta}
\end{gather*}
Except the Ricci scalar they are non-linear in the second derivatives.

For $n=4$ we have 14 invariants. They are
given explicitly 1956 by G\'{e}h\'{e}niau and Debever%
\ebnercite{Geheniau}%
. In Misner, Thorne, Wheeler%
\ebnercite{Misner}
it is stated that
all of them except $R$ are non-linear and do not have Newton's gravitation as a limit.

The Ricci-tensor, i.e. the Beltrami invariant $R^{ab}\varphi_a\varphi_b$ is absent
in the box $(G=2, B=1)$ for $n=2$. Because of
$R_{ab}= {\scriptstyle\frac{1}{2}} R g_{ab}$, it can be constructed
by lower invariants.

\subsection{\bf Was Hilbert's theory not fully covariant?}
Some historians claim Hilbert, while writing his page proofs or sending
his Nov 16, 1915, postcard,
cannot possibly have known
the correct field equations of General Relativity, since in his Axiom III of the
page proofs he defines special coordinates, which he calls space-time
coordinates, thus he had not yet obtained full covariance.
These historians overlook, what Hilbert and Einstein (in his Nov 4 paper)
have clarified almost 90 years earlier: A generally covariant theory must be
supplemented by non-covariant gauge-conditions, i.e. by specializing the
coordinates to some degree. A generally covariant theory means that I can
choose the coordinates completely arbitrary, also in the future. No
physical theory can predict my choice of the coordinates in the future
(`lack of causality'). Therefore, I must tell my choice for the future, and
this is done by imposing non-covariant so called gauge-condition,
additionally to the field equations. Such gauge conditions
should not restrict the generality of the gravitational field (a restriction
which is the duty and the right of the field equations only), i.e. the
gauge-condition must be a {\em permitted} gauge condition. To prove that a
gauge-condition is permitted, one has to show that in an arbitrary
gravitational field expressed with arbitrary coordinates, one can always transform to
new coordinates fulfilling the gauge conditions.

Hilbert's space-time coordinates are a gauge condition, but not a permitted one.
He recognized that. Therefore, he suppressed it in the published version of the page proofs.
This, however, by no means affects the general covariance of his field equations.

\subsection{\bf Are the theories of Hilbert and Einstein different?}
In his second Nov 13 postcard to Einstein, Hilbert's claims his theory to
be completely different from Einstein's. However, this remark refers
to his treatment of the energy concept of the gravitational field only.
Concerning the field equations, Einstein assumes a general
(i.e. unspecified) energy-momentum-tensor $T_{ab}$ while Hilbert
assumes the sources of the gravitational field to be expressible by
a Lagrangian density $L$, see \ebeq{6}. While deriving the gravitational field
equations, Hilbert assumes that $L$ depend only on the
$g^{\mu\nu}$ (not its derivatives) and on some variables $q_s$
and its first derivatives. Even in contemporary physics, there is no viable
(i.e. non-speculative) model of matter contradicting these
assumptions. Therefore, Hilbert's approach (as is Einstein's) is completely
general with respect to the sources of the gravitational field.
It is true, Hilbert calls the $q_s$ `the four electromagnetic potentials'
and later he specializes $L$ to that one of Mie's theory, into which
nowadays nobody believes. However, these verbal denotations and later
specializations do not affect Hilbert's derivation of the field
equations, being general with respect to the sources. Hilbert's approach
was much more far-seeing than Einstein's, since Hilbert postulates
a unified Lagrangian formulation of the whole of physics (which at that time
was gravitation and electromagnetism only), so he must be considered the
father of the concept of a unified field theory including gravitation,
to which Einstein turns only much later.

\subsection{\bf Has Hilbert found General Relativity before Einstein?}
Definitely Einstein had the idea to interpret gravitation as geometry
of 4-dimensional non-Euclidean space, Einstein had the
almost correct \ebeq{2} field equations of general relativity, and he
had pursued the theory consistently during eight years. The question
could only be if Hilbert had found the finally correct field
equations \ebeq{7} before Einstein. We have given stringent arguments for
this question to be answered affirmative. However, we have also
argued it was by Hilbert's good luck that he has achieved this. His derivation
was based on the fact stated in \ebeq{1}, which is incorrect as it stands, or
at least, when modified by the requirement of linearity in the second derivatives, was
not proved in 1915.
Hilbert has displayed an astonishing mathematical
intuition by formulating his famous 23 problems. All of them have
tuned out to be non-trivial, and most of them solvable. This was a remarkable
achievement. So also in this case he was guided by his intuition,
which in the end was not misleading.

From the point of view of physics, Hilbert did not bother about the question
if his field equations have the correct Newtonian limit. So again it was
good luck, that his choice of $R$ as the Lagrangian turned out to be the
physically correct one.

It seems that this was also Hilbert's own judgment: It is reported%
\ebnercite{ConstanceReid}
Hilbert had joked: ``Every boy in the streets of G\"ottingen
understands more about four-dimensional geometry than Einstein.
Yet, in spite of that, Einstein did the work and not the
mathematicians."

Laws of nature cannot be deduced or derived. They can only be guessed, hoping that
future experiments would verify or at least not falsify them.
This is because nature did not anticipate our wishes.

Hilbert was the first to have guessed the correct laws of gravitation, and
had communicated them on his Nov 16 postcard, which was lost or was
intentionally abolished.

\section{Mutilation of Hilbert's Page Proofs}
Since recently, very accurate photos of Hilbert's page proofs are
available on the internet%
\ebnercite{PageProofsOnTheInternet}%
. They can also be ordered as a copy from the G\"ottingen archive%
\ebnercite{OrderFromArchive}%
.

About one third of a doubly printed sheet (top of
page 7 and 8) are cut-out along a wavy line, so it was not done by scissors
but by a razor blade or a knife. Since on page 7 the cut goes irregularly
middle of a printed line, whereas on page 8 the cut goes, though wavy, but exactly
between two lines of text, it is clear the cutter intended page 8.
As judged by context and by comparing it with the published version,
the cut-off contained the definition of the gravitational part of Hilbert's
Lagrangian density, which he denoted by $K$. In the remaining part of the page
proofs it can no longer be seen, that Hilbert has chosen $K$ to be the
Ricci scalar $R$.

The page proofs have been investigated thoroughly by the G\"ottingen
historian Daniela Wuensch, who has published the results in her
book%
\ebnercite{Wuensch}
 ``Zwei wirkliche Kerle"
In the following we give a reformulation of her proof of circumstantial
evidence that Hilbert's page proofs
have been mutilated in recent years. The logic of the proof is as
follows: Several curious facts can be seen today on the page proofs
which cannot be explained in a rightful way, but which find a
plausible explanation by the motivation of a historian intending
to withdraw Hilbert's priority and the ensuing need to cover-up.

\subsection{\bf Reasons why the cut-off was done between 1994-1998}
\vspace*{0.4cm}
1) The cut-off was mentioned for the first time in T. Sauer%
\ebnercite{Sauer1}%
, an article which has appeared in 1999.

2) In 1998 the page proofs have been filmed by the G\"ottingen
archive, and on the film the cut-off is already visible.
Thus the cut-off was done before 1998.

3) The main statement in the {\em Science} 1997 article%
\ebnercite{Science}
was that in Hilbert's page proofs
the explicit field equations cannot be found. It is improbable the editors and referees
of {\em Science} had not required a copy of the page proofs, and they had remarked the cut-off
and had insisted on mentioning it. However, it is possible that personal relationships
had exceptionally made a short-cut of the usual refereeing procedures. An investigation
at {\em Science} should be initiated. Perhaps a referee or the editorial board is still
in possession of an intact copy of the page proofs.

4) In this article
a very minute comparison is made
between the page proofs%
\ebnercite{PageProofs}%
,
the published version%
\ebnercite{HilbertPublished}
and the republished version%
\ebnercite{Hilbert1924}
of 1924. In particular on p. 1272
middle column line 5, the authors of the article%
\ebnercite{Science}
write:\\
{\em
``In the proofs of his first communication, Hilbert's world function
includes a gravitational term $\sqrt{g}K$ ..."}.\\
However, they could not see that, because on the incomplete page proofs
it cannot be found, and as judged by the published version, it was
stated on the cut-off.

5) The authors would have mentioned the cut off. For historians it would be an opportunity
to speculate and to publish about such a peculiarity in a historical document.
On the contrary they let the priority of its discovery to others.

6) Not mentioning the cut-off in an article, stating the whole historical
document does not contain the field equations, would be equivalent
to a grossly unethical scientific behavior which would not be
committed without a strong motivation.

Thus CRS have been in possession of an intact copy of the page proofs.
Since they have discovered the page proofs shortly after 1994, the
cut-off was not yet done before 1994.

\subsection{\bf Reasons why Hilbert did not do the cut-off himself}
Besides the evidence just given, it is preposterous to assume Hilbert himself could have
done it. Out of over 62 publications, just this once he has kept
the page proofs, in all other cases the reprints of the published
versions only. He had namely omitted something in the final
publication because it seemed insecure to him and he intended to
improve it later.

Fetching razor blade and glue, making the cut-off and inserting it somewhere else
would cost more time than copying some simple formulae.

In 1918 Hilbert sends the page proofs%
\ebnercite{PageProofs}
to Felix Klein, who was interested to see exactly what Hilbert has
omitted in the published version%
\ebnercite{HilbertPublished}%
. In the accompanying letter%
\ebnercite{Sorgetragen}%
, dated Mar 7, 1918,
Hilbert asks Klein expressively to be careful with the page
proofs and to send them back, because he had no other records. But
the forger has overlooked that on the backside of the doubly
printed cut-off was exactly what Hilbert wanted to show Klein.

\subsection{\bf Ripping apart of a sheet and hand-foldings}
Besides the cut-off, Hilbert's page proofs have suffered
additional disfigurements. Sheet one of the page
proofs have been torn into two pieces. There is no reasonable
explanation for this fact, when everything had gone the right way.
However there is a plausible explanation for a forger by his
motivation for cover-up.

Originally the page proofs consisted of three big 4-page
sheets and of one final small sheet. The first big sheet
(frontside-backside, i.e. doubly printed) contained pages 1-2 and
beneath pages 7-8, so that after sewing, all pages come into the
correct order. (On the small sheet was page 13, backside empty.)
The big sheets have been machine-folded vertically in the middle
by the printery. Up to now, the third sheet has, beside the
machine folding, no additional (i.e. hand-) folding. Thus we can conclude the
printery has sent (Dec 6, 1915) the page proofs in a big envelope
(approx. 24 cm x 17 cm).

After the cut-out, as the first additional action, sheet one was ripped-apart
along the middle vertical machine-folding (i.e. page 1-2 was separated from
page 7-8, the latter with the cut-off).

Second, page 1-2, page 7-8
and sheet two have been folded by hand to the format 19 cm x 15 cm.

The temporal sequence of these manipulations is undoubted:
First the cut-off (almost one third of the page 7-8), then
ripping apart of sheet one (ripping off the remaining part
of page 7-8 from page 1-2), then folding to the smaller format
(19 cm x 15 cm). The cut-off was made by a razor blade
(as can be seen by the wavy cutting curve), the separation was
made by simple tearing off by hand along the machine-folding.
The cut-off has two razor blade edges (one vertical, one horizontal),
but no (vertical) tearing edge. Thus, first cut-off, then
separation of sheet one.

Only then came the foldings, because page 1-2 and
page 7-8 (being only 2/3 of a normal page) are folded
differently, namely in their respective middles.

At Mar 7, 1918 Hilbert writes a letter to Felix Klein%
\ebnercite{Sorgetragen}
in the format 19 cm x 15 cm:

{\em ``Hereby, I send you the first proofs * (3 sheets)} [= the
page proofs%
\ebnercite{PageProofs}
discussed here] {\em of my first
communication, (* Please, kindly to return them to me,
because I have no other records) in which I just have
elaborated also what are now Runge's ideas, in particular
Theorem 1, page 6, where I have proved the divergence
property of energy. But later}
[i.e. In the published version%
\ebnercite{HilbertPublished}%
] {\em I have suppressed  the
whole thing, because it did not seem mature to me. I
would be very pleased if now a progress could be achieved."}

Theorem 1 begins on bottom of page 6 and is proved on top of page
7, i.e. at that position of the page, which is now lacking.
Hilbert does not send a reprint but the page proofs, because in
the final publication he had omitted (``suppressed") something.
Even more absurd it would be to assume Hilbert has made  the
cut-off before sending the letter to Klein, so exactly that would
be missing he intended to show Klein. At least he would have
mentioned the cut-off with regret, while mentioning the "3
sheets", he is sending.

According to the temporal sequence established above,
Hilbert did not fold the page proofs to the format of the letter
(19 cm x 15 cm). Hilbert has sent the page proofs unfolded in a
big envelope (24 cm x 17 cm, perhaps in the same he had received
from the printery). The accompanying letter in the smaller format
(19 cm x 15 cm) he had simply enclosed.

{\footnotesize Is it possible Hilbert had sent the page proofs to
someone else (after Klein) and in order to save postage, he had
folded the page proofs? To everyone else, Hilbert would have sent
a reprint of the final published version, except the addressee, as
Klein, is exactly interested what Hilbert had suppressed in the
final version, namely what is written on the cut-off page 7. So
the same arguments as for Klein are valid also.
}

Now, we already have three mysteries: Who (Hilbert,
his wife K\"athe, a maid, an assistant) has performed (and why ?)
the cut-off, has ripped apart sheet one, and has folded the
pieces to the format 19 cm x 15 cm?
The cut-off, the separation of sheet one and the hand-foldings
are a fact, which cannot find a reasonable explanation when everything
has gone the right way.
However, there is a
plausible explanation for a forger.

The forger, after having done the cut-off has
pondered how he could further manipulate the page proofs in order
to prove the cut-off were done in historical times, namely before
Mar 7, 1918 (Hilbert's letter to Klein%
\ebnercite{Sorgetragen}%
).
The forger must have been in possession
of  relevant historical details. He must have known Hilbert
has sent a letter to Klein in the format 19 cm x 15 cm,
and he must have known the contents of that letter.

Obviously, the forger did not pay attention to the
argument above (about what Hilbert wanted to show Klein).
But he has taken care of for a proof of the
temporal sequence: cutting-off, ripping-apart, folding.
It has seemed plausible to the forger, Hilbert has folded  the
page proofs to the smaller format of the accompanying letter,
perhaps to save postage. Thus, seemingly, the forger could prove
the cut-off was done before Mar 7, 1918, because the foldings
presuppose the cut-off.

Notice, the forger had to rip-apart sheet one, because only now, he
could fold the mutilated page 7-8 (with the cut-off)
in the middle of the remaining 2/3 of the page, whereby
he could make sure, the foldings have taken place
before the sending of the letter (Mar 7, 1918).

\subsection{\bf The Roman numerals}

There is a further mystery: On the page proofs there
are 3 Roman numerals, not in the handwritings of Hilbert nor of
his wive K\"athe. Furthermore, Hilbert practically never used
Roman numerals. If he had to renumber, he used Arabic numerals in
a different color, Roman numerals for numerating volumes, chapters
or theorems only.

In the accompanying letter to Klein, Hilbert has written%
\ebnercite{Sorgetragen}%
:
{\em ``Hereby, I send you the first proofs * (3 sheets) of
my first communication, (* Please, kindly to return
them to me, because I have no other records.)"}

Knowing this passage, the forger had the problem
what Hilbert could have meant with ``3 sheets". Even
to him, it seemed incredible, Hilbert would have denoted
page 1-2, the large sheet two (i.e. a double page with pages 6-3
and 4-5) and the mutilated page 7-8, though all three
have a different size, collectively as ``sheets".

Therefore, the forger added Roman numerals. The Roman
number III is on page 7 directly beneath the cut-off.
(The printed number 7 had disappeared because of the
cut-off.) Thus, seemingly,  he could prove the Roman
numbers have been written after the cut-off, or in
other words the cut-off was done before posting
the letter (Mar 7, 1918).

The forger, though very cunning without any doubt,
here again has betrayed himself: Even while assuming
(already excluded above) Hilbert had sent Klein the
above mentioned variously sized ``3 sheets"
(i.e. pages 1 to 8), it is still completely incomprehensible
Hilbert would have added these Roman numerals. The consignment
already had the original  printed pagination, except as
usual the title page 1 and the mutilated page 7-8. Hilbert
simply had written a 7 and 8 beneath the cut-off. The Roman
numerals make sense only because the forger was uncomfortable
selling such different formats as the ``3 sheets" mentioned
by Hilbert.

The Roman numerals are a fact which does not have a
reasonable explanation when everything has gone the right way.
However, they find a reasonable explanation by the motivation
of a forger.

\subsection{\bf Crossing out pages 7 and 8 with a pencil}

Now we arrive at a further mystery:
The remaining 2/3 of the page 7-8 (i.e. later than the cut-off)
have been crossed through from top left to bottom right
with a pencil, again erased later on, but still slightly
visible%
\ebnercite{Bartell}%
.

These pencil vestiges are a fact, which do not have a reasonable
explanation if everything has gone the right way.
Plausible explanation for a forger: Soon after the cut-off (i.e.
before the foldings) the forger had crossed out the remaining
page 7-8, suggesting Hilbert is considering the page
incorrect, making it plausible he had cut off something
to save a bit of time. Later the forger had a better idea
of the foldings (because they would date the cut-off
before Mar 7, 1918). Now the cross out is obsolete, since
Hilbert would not cross out what he sends Klein. Also
the forger might have realized Hilbert would have
crossed out the whole pages 7-8 to mark them as obsolete,
and eventually later on he had cut-off something from
these wrong pages in order to save time. After the
cut-off the pages were sufficiently marked as irrelevant
and there were no need of an additional crossing-out.

\subsection{\bf Wrong Archive-pagination}
One more mystery: Old handwritings or
valuable printed documents (such as Hilbert's page proofs) coming
new into the G\"ottingen archive, are paginated by an archivist in
the middle of the right edge using a pencil. In the following we
will call it `Archive-pagination' or `A-pagination' for
short. As a rule, this A-pagination is done with great care.
Already originally  (i.e. print) paginated documents (such as
Hilbert's page proofs, pages 1-13), as a rule, are not paginated
again.

It is a mystery Hilbert's page proofs have been
paginated at all, and in a completely wrong order, especially
astonishing since an original printed pagination was present. The
A-pagination on the mutilated page was done at the right edge
exactly in the middle of the remaining 2/3 of the page, making it
clear - apparently - the A-pagination was done after the cut-off.

In 1997 Corry phones the archive, reporting the
completely wrong A-pagination (wrong order). The
archive rubs out the old (wrong) A-pagination and
replaces it by a correct A-pagination. With a
magnifying glass the old A-pagination can still be
recognized.

The mystery of the wrong A-pagination finds a plausible
explanation by supposing the page proofs have not been paginated
while they arrived into the archive, since they had already the
original printed pagination. After the forger had done the
cut-off, had ripped-apart the second sheet, he added an
A-pagination. This should prove the page proofs already had the
cut-off while they arrived into the archive. The intentional wrong
A-pagination makes sense, because a new A-pagination must be done,
so the handwriting of the forger disappears.

\subsection{\bf Forgery of a date and a dog-ear in Hilbert's private notes}
As we have seen in subsection \ebeq{6.2} the sheet with archive page number 32
in Hilbert's folder ``{\em {\"U}ber Elektrodynamik}" are his notes while he
has found the correct coefficients
${\ebnerboldmath \alpha}=1$, ${\ebnerboldmath \beta}=-1/2$ and
${\ebnerboldmath \lambda}=0$ in the explicit
field equations. As can be seen on a facsimile in Wuensch%
\ebnercite{Wuensch}
p. 60, this sheet
32 on top left bears the date 9/10.IX.15. It is certainly a good habit
to add a date in all our own private notices. But had Hilbert had that habit? At no
other sheet of his folder we can find such a date.  Only with poor
graphological capabilities one would suspect that the date has not
the same handwriting as the remainder of the page, especially since both
should be written in the same minute, with the same quill and with the same ink,
and especially comparing it with the facsimile (Wuensch, p. 22) of a date
on Hilbert's letter to Klein. A possible motivation for a forger would be
to prove Hilbert has found or verified the explicit field equations only after he has
seen them in Einstein's publications.

But the curiosities continue. I have ordered a copy of Hilbert's folder
from the G\"ottingen archive%
\ebnercite{OrderFromArchive}
and I was struck the date is now absent. From G\"ottingen I
got the answer a turned-down-corner (dog ear) at the top left of Hilbert's page 32 sheet
has prevented the date to become visible on the microfilm from which all
requested copies are now made. Coincidence or an attempt by anyone who is
interested to reduce the number of too obvious and thus counterproductive forgeries?

\subsection{\bf How is it possible to manipulate documents in the G\"ottingen archive?}
For the question how it is possible to smuggle documents
out of the G\"ottingen archive and to manipulate them alone,
I refer to Wuensch's book%
\ebnercite{Wuensch}%
.
For instance it is possible to
put them between one's own sheets, because while leaving
the archive for a rest, personal working material is not
checked.

\subsection{\bf What was on the cut-off in Hilbert's page proofs}

There is agreement among historians the cut-off contained equation
\begin{equation*}
\tag{41}
H = K + L
\end{equation*}
where $H$ is the total Lagrangian density, and $K$ was defined on the cut-off e.g. as
`the invariant stemming from Riemann's tensor', i.e. $K \equiv R =$ Ricci-scalar.

Some historians believe the cut-off also contained the explicit field equations \ebeq{5c}.
Though we cannot exclude this possibility definitely,
there are several arguments (see also T. Sauer%
\ebnercite{Sauer2}%
) that this is improbable:

The cut-off is rather short, for
all the definitions and explanations. Furthermore, the subject just before the cut-off
is different, thus some transitional phrasing would be required. To give the
explicit field equations does not fit to the position of the cut-off but
rather after Eq. (26) of the page proofs,
a position where they appeared in the printed version.

In the whole article (page proofs, published version and
republished version) Hilbert was mainly interested in general
theorems and in the energy concept. Hilbert did not include
explicit formulae for the electromagnetic field equations either.

The most decisive argument, however, is the following: the editors of {\em Science}
would not have passed the 1997-article without seeing a copy of the page proofs.
They had observed the cut-off, since the main statement of the paper is, that nowhere
in the whole page proofs the explicit field equations can be found. Thus, we can
conclude that at the discovery of the page proofs in the G\"ottingen
archive short after 1994, the cut-off was not yet there. Seeing the complete page-proofs,
the editors had recognized the explicit field equations, even if formulated in a
version like \ebeq{5c}, different from a modern one. Thus we can conclude the
cut-off did not contain the explicit field equations.

\subsection{\bf What was the motivation of the forger?}
Some historians believe the cut-off on Hilbert's page proofs contained
the explicit field equations. Then the motivation would be to withdraw
Hilbert's priority to have found the explicit field equations before Einstein,
and to be able to publish this new historical perspective.

In the above subsection we have collected arguments
the cut-off did not contain the explicit field equations.
So what could then be the motivation for a forger to perpetrate the cut-off?

One strange uttered hypothesis%
\ebnercite{Renn}
is that self appointed advocates of Hilbert
had done the cut-off in order at least to have a chance to speculate Hilbert
has found the field equations before Einstein.

But there is a more serious and very plausible motivation:
For a historian of science, it is a great triumph to be able to reverse the established
opinion about the priority of a milestone in scientific progress, as was the
final formulation of general relativity.
After having detected the page proofs did not contain the explicit field equations,
this success of that historian was jeopardized by the argument Hilbert was
nevertheless the first to have given
the Ricci-scalar as the correct Lagrangian density for the gravitational field, and
the evaluation of the corresponding explicit field equations was merely a
straightforward technical exercise.

The human psychology reacts more to a fast change of possession
than to its actual level. Thus in a fit of ill-considered activity the historian,
or someone else enjoyed by that initial success,
made the cut-off
by removing with a razor blade Hilbert's definition of $K$ as the Ricci-scalar.
Thus he hoped to argue that Hilbert has made his choice for $K$ only
while preparing for his printed version and he has added the definition of
$K$ as the Ricci-scalar only then, together with the corresponding explicit
field equations. And that Hilbert had made the cut-off himself in order to paste some
formulae somewhere else in order to save time.

At no other place on the remaining page proofs it can be decided if Hilbert had
really taken the Ricci scalar, or if he had left his Lagrangian density unspecified.
The only remaining hint is Hilbert's choice of the letter $K$, which is similar
to $k$ which was used by Gauss for curvature, see \ebeq{11}. Even that is now questioned by
T. Sauer%
\ebnercite{Sauer2}%
,
footnote 7, arguing Hilbert has taken $K$ for alphabetical reasons.
Except that hint, thus far, nobody doubts Hilbert had indeed defined on the cut-off
his Lagrangian density $K$ as `the invariant stemming from Riemann's tensor', i.e the
Ricci scalar, as can be seen by the printed version.
However, we are not in need to argue the forger was very anticipating.

\subsection{\bf Summary: a chronology of the forgery}
After discovery of Hilberts intact page proofs in a G\"ottingen
archive after 1994, it was observed they did not contain the
correct explicit field equations of General Relativity. A
publication of this spectacular historical fact, reversing the
priority dispute in favor of Einstein, was begun.

Shortly, by internal discussions, it was argued Hilbert still
deserves the priority
of having found the correct Lagrangian density, and the
evaluation of the explicit field equations could be considered
a technical computational exercise only.

Therefore, a forger, who must have detailed insight in the relevant physics
and mathematics, made the shortest possible cut-off, so that to day it is
no longer possible to prove with the remaining part of the page proofs
Hilbert had taken the Ricci scalar as the Lagrangian density.
One could now argue Hilbert had
inserted not only the correct field equations but also the
correct Lagrangian density only after Dec 6, 1915.

The forger, possibly plagued by remorse and anxiety, thought about
further actions with the intention to prove the cut-off was done in
Hilbert's time. First with a pencil, he crossed out the remaining two-thirds
of pages 7 and 8, what he rubbed out later.
These traces are a fact on the page proofs, which do not have a reasonable
explanation if everything had gone the right way. However, they find
a plausible explanation by the motivation of a forger. One should assume
Hilbert had made the crossings indicating the contents of the
pages are obsolete. So it becomes more probable Hilbert has made the cut-off himself
to save time. Later, the forger recognized Hilbert would have done the crossing
before the cut-off. After the cut-off the sheet was already sufficiently marked
as unimportant.
Therefore, the forger erased them again, removing this nonsensical action.

Only now the forger was  thinking more seriously about decisive
manipulations in order
to prove Hilbert has done the cut-off himself. The forger must be in possession of
relevant historical data, and in particular about the format and contents
of Hilbert's letter to Klein%
\ebnercite{Sorgetragen}%
. He asked himself: ``How could I prove Hilbert
has made the cut-off before sending the page proofs to Klein?".
His answer: ``I could fold the page proofs to the format of the accompanying letter
to Klein, and if I fold the mutilated page in the remaining middle, I can
prove the cut-off was done before sending the letter, i.e. before 1918."
But the forger had a problem, because pages 1-2 (frontside-backside) are still
connected with the mutilated pages 7-8 on the same large sheet. Therefore
he separated page 1-2 from pages 7-8. Only then he was able to make the differing
hand-foldings in their
respective middle. In the vertical direction he made foldings to the
format of the accompanying letter to Klein. But now he had a further problem,
because in the accompanying letter Hilbert writes that he will send 3 sheets.
The forger had three very different sheets: half a sheet (pages 1-2),
an intact sheet (pages 3-4 and 5-6) and a mutilated sheet (pages 7-8). It seemed
strange Hilbert had called them collectively as `3 sheets'. Therefore
the forger added three Roman numerals to make that more plausible.

The forger observed Hilbert's page proofs had no archive pagination. This
gave him the opportunity to add one with the pagination in the
middle of the mutilated page. Thus he could prove the cut-off was already there
when the page proofs came into the archive, which is the time the archive
pagination is usually done.
\section{Acknowledgements}
I thank Dr. D. Wuensch for the communication of several details
not yet contained in her book%
\ebnercite{Wuensch}%
,
and also Dr. K.P. Sommer and Prof. F. Winterberg
for clarifying answers to several questions, and Prof. H. Dehnen
for a discussion about the range of empirical validity of Eq. \ebeq{2}.



\begin{thebibliography}{99}
\bibitem[36]{HilbertErroneously}
Hilbert shows in his page proofs in his Theorem I, which is
an anticipation of what is now called Noether's theorem, that,
because of general covariance,
among the 10 gravitational and 4 electromagnetic equations there
exist 4 identities, so e.g. the 4 gravitational equations are a
mathematical consequence of the 10 gravitational equations.
However, Hilbert overlooks that this is the case only, because
of the electrodynamic terms in the 10 gravitational equations,
which originate by his adding an electrodynamic term $L$ to
the Ricci scalar $K \equiv R$ in his Lagrangian density.
So Hilbert's opinion to have shown electromagnetism as a
consequence of gravitation is untenable. However,
it must be emphasized that Hilbert was perhaps the first to
have proposed a unified field theory of gravitation and
other interactions in the form of a single Lagrangian density.
\bibitem[39]{Schur}
Today F.Schur's theorem is proved using the contracted Bianchi
identities.
\end{thebibliography}
\end{document}